# CNeuroMod-THINGS, a densely-sampled fMRI dataset for visual neuroscience

**Marie St-Laurent** [1,2,✉], **Basile Pinsard** [2], **Oliver Contier** [1], **Elizabeth DuPre** [2,3], **Katja Seeliger** [1,4], **Valentina Borghesani** [2,5], **Julie A. Boyle** [2], **Lune Bellec** [2,3], **and Martin N. Hebart** [1,6,7]

[1]Max Planck Institute for Human Cognitive and Brain Sciences, Leipzig, Germany
[2]Centre de recherche de l'Institut universitaire de gériatrie de Montréal, Montréal, Canada
[3]Département de psychologie, Université de Montréal, Montréal, Canada
[4]Martin Luther University Halle-Wittenberg, Medical Faculty, Halle, Germany
[5]Faculté de psychologie et des sciences de l'éducation, Université de Genève, Genève, Swizerland
[6]Department of Medicine, Justus Liebig University Giessen, Giessen, Germany
[7]Center for Mind, Brain and Behavior (CMBB), Universities of Marburg, Giessen, and Darmstadt, Germany

**Data-hungry neuro-AI modelling requires ever larger neuroimaging datasets. CNeuroMod-THINGS meets this need by capturing neural representations for a wide set of semantic concepts using well-characterized images in a new densely-sampled, large-scale fMRI dataset. Importantly, CNeuroMod-THINGS exploits synergies between two existing projects: the THINGS initiative (THINGS) and the Courtois Project on Neural Modelling (CNeuroMod). THINGS has developed a common set of thoroughly annotated images broadly sampling natural and man-made objects which is used to acquire a growing collection of large-scale multimodal neural responses. Meanwhile, CNeuroMod is acquiring hundreds of hours of fMRI data from a core set of participants during controlled and naturalistic tasks, including visual tasks like movie watching and videogame playing. For CNeuroMod-THINGS, four CNeuroMod participants each completed 33-36 sessions of a continuous recognition paradigm using approximately 4000 images from the THINGS stimulus set spanning 720 categories. We report behavioural and neuroimaging metrics that showcase the quality of the data. By bridging together large existing resources, CNeuroMod-THINGS expands our capacity to model broad slices of the human visual experience.**

**Correspondence:** ***laurentm@cbs.mpg.de, stlaurent.marie@criugm.qc.ca***

## 1. Background & Summary

The growing availability of large neuroimaging datasets is creating new opportunities to apply data-hungry computational techniques to model how the brain supports cognitive functions like perception and object processing. We introduce CNeuroMod-THINGS, an extensively sampled functional magnetic resonance imaging (fMRI) dataset that captures brain responses across a broad segment of the human visual landscape. Four subjects from the Courtois Project on Neural Modelling (CNeuroMod) each completed between 33 and 36 fMRI sessions of a continuous image recognition task during which they were shown around 4000 naturalistic images from the THINGS dataset covering 720 categories of concrete nameable objects (1). By design, the CNeuroMod-THINGS dataset forms a bridge between two large data ecosystems, CNeuroMod and the THINGS initiative. In doing so, it expands our ability to model visual brain processes along semantically diverse dimensions defined by a well-characterized stimulus set, using subject-specific data from the most extensively scanned neuroimaging participants to date.

The CNeuroMod-THINGS dataset adds to a growing number of large fMRI datasets that also feature brain responses to naturalistic images, including BOLD5000 (2), the Natural Scenes Dataset (NSD) (3), the fMRI dataset from THINGS-data (THINGS-fMRI) (4) and the Natural Object Dataset (NOD) (5). Importantly, CNeuroMod-THINGS contributes to the growing collection of datasets assembled under the THINGS initiative, which includes multimodal behavioral, neurophysiological and neuroimaging correlates of a common core set of stimulus images (4, 6). These images provide a broad, comprehensive and systematic sampling of nameable object concepts from the American English language, in contrast with other large image datasets that focus primarily on size rather than sampling of semantic space, and that feature strong biases toward overrepresented object categories. The THINGS images are also accompanied by a growing body of meta-data, ratings and annotations (7–10), including 4.7 million human judgments of perceived image similarities collected from over 12,000 participants via online crowdsourcing (4).

For the CNeuroMod-THINGS dataset, four participants were shown images from the same 720 object categories sampled by the THINGS-fMRI dataset (4), which were selected to be visually and conceptually representative of the full THINGS image set. Each participant was shown the same set of approximately 4,300 images (6 images/category), making it possible to contrast representations across individuals. Images were shown three times per participant according to a continuous recognition paradigm adapted from the NSD task (3). The inclusion of image repetitions makes CNeuroMod-THINGS the only fMRI dataset based on THINGS that supports data-driven analyses at the single image level. For comparison, the THINGS-fMRI dataset includes twice as many unique images from the same 720 categories each shown once to three participants (4), while NSD includes ~73k unique images shown three times to at least one of eight subjects (3). Of note, NSD maximized the number of images shown by presenting mostly distinct stimuli to each of their participants. With the current paradigm, we aimed to strike a balance between wide sampling and robust image-specific signal that can be compared across individuals.

Crucially, the CNeuroMod-THINGS dataset is part of CNeuroMod (11), a deep phenotyping (12) project for which six core subjects have each completed several controlled and naturalistic fMRI tasks, including movie watching, video game playing, listening to and recalling narratives, resting state, reading, working memory and language tasks. Other deep phenotyping fMRI datasets include MyConnectome (13), the Midnight Scan Club (14) and the Individual Brain Charting (IBC) (15–17) datasets. Notably, the CNeuroMod subjects are the most extensively scanned neuroimaging participants to date, with approximately 200 hours of fMRI data per subject that include around 80 hours of video watching (18). Four of these exceptionally well-characterized individuals each completed 33-36 sessions of the CNeuroMod-THINGS task, complementing free-viewing video data with controlled image viewing defined by the THINGS stimulus set. CNeuroMod's deep phenotyping approach makes it possible to train and test models of visual brain processes under naturalistic and well-controlled conditions using data from single subjects, and to combine data across tasks that target different modalities and cognitive domains in order to build versatile individual models of brain function (12).

The core CNeuroMod-THINGS dataset includes raw and pre-processed fMRI data and key derivatives like trial-specific beta scores estimated at the voxel level. It also comprises naturalistic image stimuli and annotations that characterize their content, behavioural data that reflect performance on the image recognition task and eye-tracking data to assess trial-wise gaze fixation. To help delineate subject-specific variability, we also provide fMRI data from two vision localiser tasks—fLoc (19) and retinotopy (20)—and derivatives that include individually-defined functional regions of interest (ROIs). Finally, the current data release includes anatomical scans and whole-brain patches to project statistical results onto individual flat maps of the cortical surface. With the results reported below, we characterize the CNeuroMod-THINGS dataset and report proof-of-concept analyses that showcase the quality of the data.

## 2. Methods

### 2.1 Participants.

The Courtois Project on Neural Modelling (CNeuroMod) has acquired hundreds of hours of fMRI data from six core participants using a large variety of tasks (11). Four of the six CNeuroMod participants contributed to the CNeuroMod-THINGS dataset: sub-01, sub-02, sub-03 and sub-06. All were healthy right-handed people with no record of neurological disorders, normal hearing and normal or corrected-to-normal visual acuity for their age (aged 39 to 49 at the beginning of acquisition). Two were female (sub-03 and sub-06) and two were male (sub-01 and sub-02). Participants provided informed consent for participation and data sharing. The research was approved by the Comité d'éthique de la recherche — Vieillissement et neuroimagerie — of the CIUSSS du centre-sud-de-l'île-de-Montréal (under number CER VN 18-19-22).

### 2.2 Task stimuli.

Stimulus images were selected among 720 of the 1854 categories of images available through the THINGS initiative. The THINGS images provide a broad and systematic sampling of object concepts that is representative of the American English language, with each category depicting a unique nameable concept that is either manmade or natural (1). The 720 categories used in the current study were also used to collect the THINGS-fMRI dataset (4), and were selected to be visually and conceptually representative of the full THINGS image set. To characterize images during data driven analyses, we used higher order categorical labels (e.g., "animal", "plant") and object concept annotations (e.g., "size", "natural") generated by the THINGSplus project (9), as well as boolean flags that reflect image content (e.g., the presence of human or animal faces) generated manually by author MSL (Supplementary Table S1).

The three participants (sub-01, sub-02 and sub-03) who completed 36 sessions of the image recognition task were shown 6 unique images per category (the first 6 images of a category based on their numbering in the THINGS image set). Sub-06, who completed 33 sessions, was shown 5 images for 480 categories, and 6 images for the remaining 240. With the exception of 120 images shown for the first time during the last session, every image was repeated once within a session and once between consecutive sessions. In total, participants saw 4320 unique stimuli throughout the experiment (3840 for sub-06). By comparison, for the THINGS-fMRI dataset from THINGS-data (4), 8,640 unique images from the same 720 categories (12 images / category) were shown once (and 100 images were shown 12 times) to three participants who completed 12 fMRI sessions, sampling twice as many images as the current paradigm with a single presentation for most images.

### 2.3 Continuous image recognition task paradigm.

**Trial structure.** Participants completed between 33 (sub-06) and 36 (sub-01, sub-02 and sub-03) fMRI sessions during which they performed a continuous image recognition task (Fig. 1a) designed to ensure subject engagement without introducing block-design effects in the neural signal (3). The first session included 3 fMRI runs while all subsequent sessions included 6 runs, totaling 213 runs for sub-01, 02 and 03, and 195 runs for sub-06. Each run included 60 experimental trials during which 190 functional brain volumes (TR=1.49s) were acquired over 283s. For each trial, a single $900 \times 900$ pixel stimulus image was presented in the center of a $1280 \times 1024$ screen, occupying 10° of visual angle. The image was shown for 2.98s, followed by a 1.49s ISI (onset and offset times were time-locked to the fMRI sequence). A black fixation marker (2° of visual angle) combining cross hair and a bulls eye (a central dot and four wedges located cardinally (21)) was visible at all times and overlaid onto the image center during image presentation. Participants were instructed to maintain fixation on the central dot throughout a run.

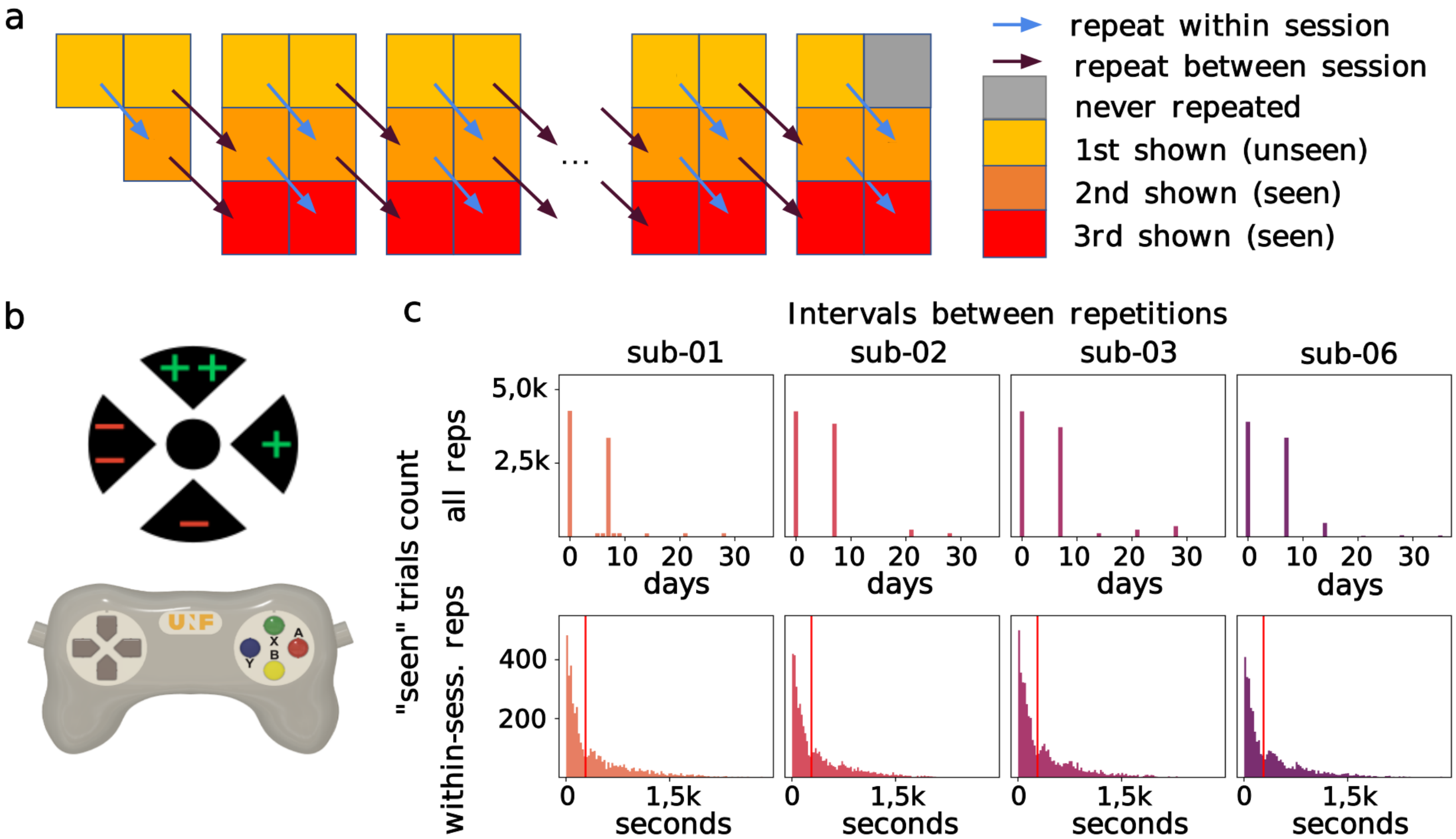

**Fig. 1. Experimental paradigm. a.** Unique stimulus images were shown three times over the course of the experiment. fMRI sessions are represented as $3 \times 2$ blocks where each square represents the proportion of images shown for the first (yellow), second (orange) and third (red) time. Half the images were first repeated within and then across sessions, and vice versa. Trials from each type (1st, 2nd and 3rd viewing, repeated within/between sessions) were intermixed within each run. **b.** Response-to-button mapping symbol shown on the central fixation marker (sure seen: $++$, unsure seen: $+$, unsure unseen: $-$, and sure unseen: $--$). Responses were made with the right thumb by pressing buttons on a custom-made MRI compatible video game controller [top: X (green), bottom: B (yellow), left: Y (blue), right: A (red)]. **c.** Frequency distribution of delays between image repetitions, per trial (2nd and 3rd viewings only), for each subject. Top charts: repetition delays for all repeated trials, measured in days. Images were either repeated within session (0 days) or between consecutive sessions, most of which were 7 days apart. Bottom charts: delays for within-session repetitions only (2nd and 3rd viewing), measured in seconds. The red vertical line indicates the duration of a single run (283s), illustrating how the majority of within-session repeats were within the same run.

**Subject response.** Most stimulus images were presented three times throughout the duration of the experiment to capture robust image-specific responses. For each trial, participants reported whether the displayed image was shown for the first time ("unseen") or whether it had been shown previously ("seen"), either during the current or a previous session (or both). Participants also reported whether or not they felt confident in their answer. Responses (seen/unseen $\times$ sure/unsure) were made with the right thumb by pressing one of four buttons (top, bottom, left, right) on a video game controller (Fig. 1b). The response-to-button pairing was indicated on the fixation marker wedges during image presentation ($++$ sure seen, $+$ unsure seen, $-$ unsure unseen, $--$ sure unseen). To dissociate memory responses from motor responses, the response-to-button pairing varied from trial to trial with random vertical and horizontal flips. Participants could provide their answer until the next image appeared. Multiple answers were recorded within the response time window to allow for self-correction. Results reported below are based on the first recorded button press. Behavioural metrics derived from the last button press and raw records of all button presses are also included in the released dataset. No feedback was given to participants about their performance on the memory task throughout the entire duration of the longitudinal experiment.

**Image repetition pattern.** A given image was either repeated first within-session and then between-sessions, or vice versa. Runs from the first session (3 runs) had a 2:1 ratio of unseen and seen images, and no between-session repeats. Runs from subsequent sessions (6 runs per session) had a 1:2 ratio of unseen and seen images, and an equal number of images shown for the first, second and third time within each run (20 each). The number of within- and between-session repeats were the same for all sessions (except for session 1), while the ratio of within-to-between image repeats increased over runs within the course of a session. Typically, the within-to-between repeat ratio was 12:28, 16:24, 20:20, 20:20, 24:16 and 28:12 for runs 1 to 6, although a small number of sessions administered out of the pre-planned order featured atypical patterns of repetition (see Supplementary S1 for a list of misordered sessions). Within a typical run, half the within- and half the between-session repeats were for second and third time repeats, respectively. Most sessions were interspaced by one week, with a few exceptions due to scanner and participant availability (Fig. 1c).

**Image categories.** No more than one image per category (out of 720) was shown per session in order to minimize interference. For each subject, image categories were randomly assigned to one of six folds of 120 categories. Each fold was used in systematic rotation to determine the categories of the novel images introduced in a session. One half of these novel images were randomly set to be repeated within the current and then the subsequent session; the other half was set to be repeated twice during the subsequent session. The exemplar image from each category was determined with an order unique to each subject (e.g., 2, 4, 3, 6, 5, 1), so that all exemplars with a given number (here starting with 2; e.g., alligator_02s.jpg, banner_02s.jpg) were introduced during a block of 6 consecutive sessions, followed by exemplars with the next number (e.g., alligator_04s.jpg) in the next 6 sessions, etc. Stimulus images were assigned to runs by sampling randomly among the novel and repeated image viewings planned for a given session according to the number of trials per subcondition planned for each run. For example, for a first run, 6 and 14 images were sampled randomly among the novel images set to be repeated within and between sessions, respectively (totalling 20 novel image trials in that run). The order of trial presentations was also randomized within each run. The exact stimulus ordering can be re-created using the ses-thingsmem.py script.

## 2.4 MRI setup and data acquisition.

Participants were scanned with a Siemens PRISMA Fit scanner equipped with a 64-channel receive head/neck coil available at the functional neuroimaging unit (UNF) in the Centre de Recherche de l'Institut Universitaire de Gériatrie de Montréal (CRIUGM). During scanning, each participant wore a personalized polystyrene headcase to minimize head movement (22). Visual stimuli were projected with a Epson Powerlite L615U projector onto a blank screen positioned behind the MRI bore made visible to the participant through a mirror mounted on the head coil. The presentation of stimuli, the recording of responses and the synchronization of the task with scanner trigger pulses were performed with a custom overlay based on the PsychoPy library ( $>=$ v2020.2.4) (23). This software also triggered the onset and calibration of the eye-tracking system, which collected eye-tracking data from the right eye at 250Hz with Pupil Core software (24) and a head-coil mounted MRC High-Speed camera. Participant responses to stimuli were collected using a 3D printed custom-made MRI compatible video game controller (25).

Throughout each session, physiological data were acquired using BIOPAC MP160 MRI compatible systems and amplifiers (BIOPAC AcqKnowledge 5.0 software, 10000 Hz sampling rate). Physiological signals included electrocardiogram (ECG) activity (EL-508 wet electrodes, ECG100C-MRI amplifier), plethysmography (PPG; TSD200-MRI transducer, PPG100C-MRI amplifier), skin conductance (EDA; EL509 dry electrodes with BIOPAC 101A isotonic gel, EDA 100C-MRI amplifier) and respiratory activity (DA100C amplifier with a respiratory belt, TSD221-MRI transducer). All data acquisition scripts are available in this CNeuroMod repository: https://github.com/courtois-neuromod/task_stimuli.

Task sessions included only functional MRI runs. fMRI data were acquired with an accelerated simultaneous multi-slice, gradient echo-planar imaging sequence (26) developed for the Human Connectome Project (27) (slice acceleration factor = 4, TR = 1.49s, TE = 37 ms, flip angle = 52°, 2mm isotropic spatial resolution, 60 slices, $96 \times 96$ acquisition matrix). All fMRI data were preprocessed using the fMRIprep pipeline (28, 29) (version 20.2.5; slice-timing correction was applied). For the data quality analyses reported below, all images were processed in native subject space (T1w). Preprocessed BOLD data are available in both T1w and MNI space in the full data release. Anatomical images of each participant were acquired periodically during separate dedicated sessions (30). Structural data were acquired using a T1-weighted MPRAGE 3D sagittal sequence (duration = 6:38 min, TR = 2.4 s, TE = 2.2 ms, flip angle = 8° , voxel size = 0.8 mm isotropic, R=2 acceleration) and a T2-weighted FSE (SPACE) 3D sagittal sequence (duration = 5:57 min, TR = 3.2 s, TE = 563 ms, voxel size = 0.8 mm isotropic, R=2 acceleration). The T1w and T2w scans from each subject's first two anatomical sessions were coregistered, averaged and preprocessed with sMRIprep version 0.7.0 (31). Please see the Courtois-Neuromod documentation for additional details on data acquisition and preprocessing.

## 2.5 Cortical flat maps.

Brain surfaces reconstructed using recon-all (FreeSurfer 6.0.1, RRID:SCR_001847, @fs_reconall) by the sMRIprep pipeline version 0.7.0 were cut manually into whole-brain cortical patches and flattened with TkSurfer 6.0.0 to produce individual cortical flat maps. Flat maps were imported into PyCortex 1.2.5 (32) to support the visualisation of brain data onto individual surfaces. The current release includes cortical flat maps that can be used to visualize results from any CNeuroMod dataset for all six CNeuroMod subjects, including the four subjects who completed the CNeuroMod-THINGS task.

## 2.6 Fixation Compliance.

We derived trial-wise measures of fixation compliance from in-scan eye-tracking data. We performed quality checks to exclude runs with missing, corrupt or low quality (i.e., very noisy) data, and performed drift-correction on the remaining runs with the following steps. First, pupils detected with low confidence by Pupil Core software were filtered out (threshold > 0.9 for sub-02, sub-03, and sub-06; a lower threshold of > 0.75 was adopted for sub-01 because Pupil Core had more difficulty detecting that participant's pupil). Then, gaze positions recorded during a trial were realigned (i.e., drift corrected) based on their distance from the median gaze position during the last known period of central fixation (i.e., the reference period), which we assumed to correspond closely to the central fixation mark. Because sustained central fixation was required throughout the task, this reference period was defined as the previous trial's image presentation and subsequent ISI. For a run's first trial, its own image presentation and ISI was used as the reference period.

To estimate fixation compliance, we calculated the proportion of drift-corrected gaze points within different bins of visual angle from central fixation (0.5, 1.0, 2.0 and 3.0 ° ) during the image viewing portion of each trial. We also compiled trial-wise quality metrics like captured gaze count (to flag eye-tracking camera freezes) and the proportion of pupils detected above 0.9 and 0.75 confidence thresholds (to estimate data quality; e.g., good camera focus). To estimate gradual shifts in head position, we also calculated the distance in median gaze position during image viewing between consecutive trials. These trial-wise metrics are included in the *events.tsv file for each run for which usable eye-tracking data were available.

### 2.7 Single trial response estimate with GLMsingle.

We estimated single trial responses to individual image presentations with beta scores computed with the GLMsingle toolbox (33). BOLD volumes in native (T1w) functional space preprocessed with fMRIPrep were masked with a whole brain functional mask and normalized (i.e., z-scored) within voxel along the time dimension. The first two volumes of each run were dropped for signal equilibrium before submitting BOLD data to GLMSingle (https://github.com/cvnlab/GLMsingle.git at commit c4e298e). Denoising was performed internally by the GLMsingle toolbox with GLMdenoise (34). Cross-validation was performed to select denoising and regularization (ridge regression) parameters to prevent overfitting and improve the amount of variance explained by the beta scores. We specified a custom 13-fold cross-validation scheme (15-17 runs per fold) for which consecutive runs (e.g. run 6 of session 4, followed by run 1 of session 5) were systematically assigned to the next fold, so that runs from two consecutive sessions were never assigned to the same fold. In this manner, trials with the same image were split across at least two folds, since images were repeated once between and once within sessions (sometimes but not always within the same run; Fig. 1c). fMRI sessions were specified to the model to account for gross changes in betas across sessions during hyperparameter selection. Final trial-wise beta scores were estimated with the best combination of hyperparameters selected for each voxel.

The final beta scores were normalized (z-scored) across the entire dataset (i.e., across all voxels and all trials) and saved as trial-specific beta maps (one map per trial). Beta maps associated with the same stimulus image were also averaged (over 1-3 repetitions) and saved as image-specific beta maps. All maps are included in the data release.

### 2.8 Analyses of memory conditions.

All behavioural and fMRI analyses of memory performance excluded trials with no recorded answer and trials from session 1 during which the absence of between-session repetition reduced the task difficulty. Analyses also excluded a handful of trials impacted by out-of-order sessions (see Supplementary S1. Misordered sessions) that modified the planned repetition pattern for a subset of images. Specifically, we excluded sub-03's sessions 24-26 and sub-06's sessions 19-26 from all analyses of memory performance.

To assess whether recognition effects are present in the BOLD data, we performed t-tests on normalized trial-wise beta scores estimated with GLMsingle (33) for each subject, using a procedure similar to the one adopted for the Natural Scenes Dataset (3). Specifically, we used behavioural responses to identify trials for which a subject successfully recognized previously shown images as "seen" (hits), and correctly identified never-shown images as "unseen" (correct rejections). Our task design allowed us to further dissociate hits for images last repeated within and between consecutive imaging sessions ("within-session hits" and "between-session hits", respectively), highlighting memory recognition after short and long retention intervals (between-session hits typically followed a 7-day retention interval; Fig. 1c).

For each voxel, we performed two-sample t-tests comparing betas from either "within-session hit" or "between-session hit" trials to betas from "correct rejections" trials. Betas were concatenated per subject across all sessions. Unequal variance was allowed across conditions in the two-sample t-test to account for variability in the relative difficulty of long- and short-term memory recognition compared to correct rejections. The resulting t-values are included in the released dataset.

### 2.9 Noise ceilings.

We computed voxelwise noise ceilings on trial-specific beta maps to estimate the maximal proportion of beta score variance that could be explained by the identity of the stimulus image, given the presence of measurement noise. Noise ceilings were estimated with a technique described by Allen et al (3), which assumes that voxel variance can be separated into stimulus-driven signal and unrelated noise. Raw voxel betas were split into repetitions 1, 2 and 3, and standardized (z-score) across images within each repetition. A voxel's noise variance was estimated as the beta variance across repetitions for a given image (normalized with $n-1$ to correct for small sample size), averaged across all images. As the variance of the standardized betas is 1, we estimate the signal variance as (1 - the noise variance), corrected with a half-wave rectification. We then calculate the noise ceiling as:

$$(100 \times ncsnr^2)/(ncsnr^2 + 1/n),$$

where $n = 3$ is the number of repetitions per image, and $ncsnr$ is the signal standard deviation divided by the noise standard deviation (3). We excluded images from this calculation that did not have a behavioral response (i.e., button press) on three distinct trials, with the assumption that subjects may have been inattentive during no-response trials. Final noise ceiling scores for each subject were based on 3247 (sub-01), 4178 (sub-02), 4179 (sub-03) and 3398 (sub-06) images, respectively. Voxelwise noise ceilings are included in the data release.

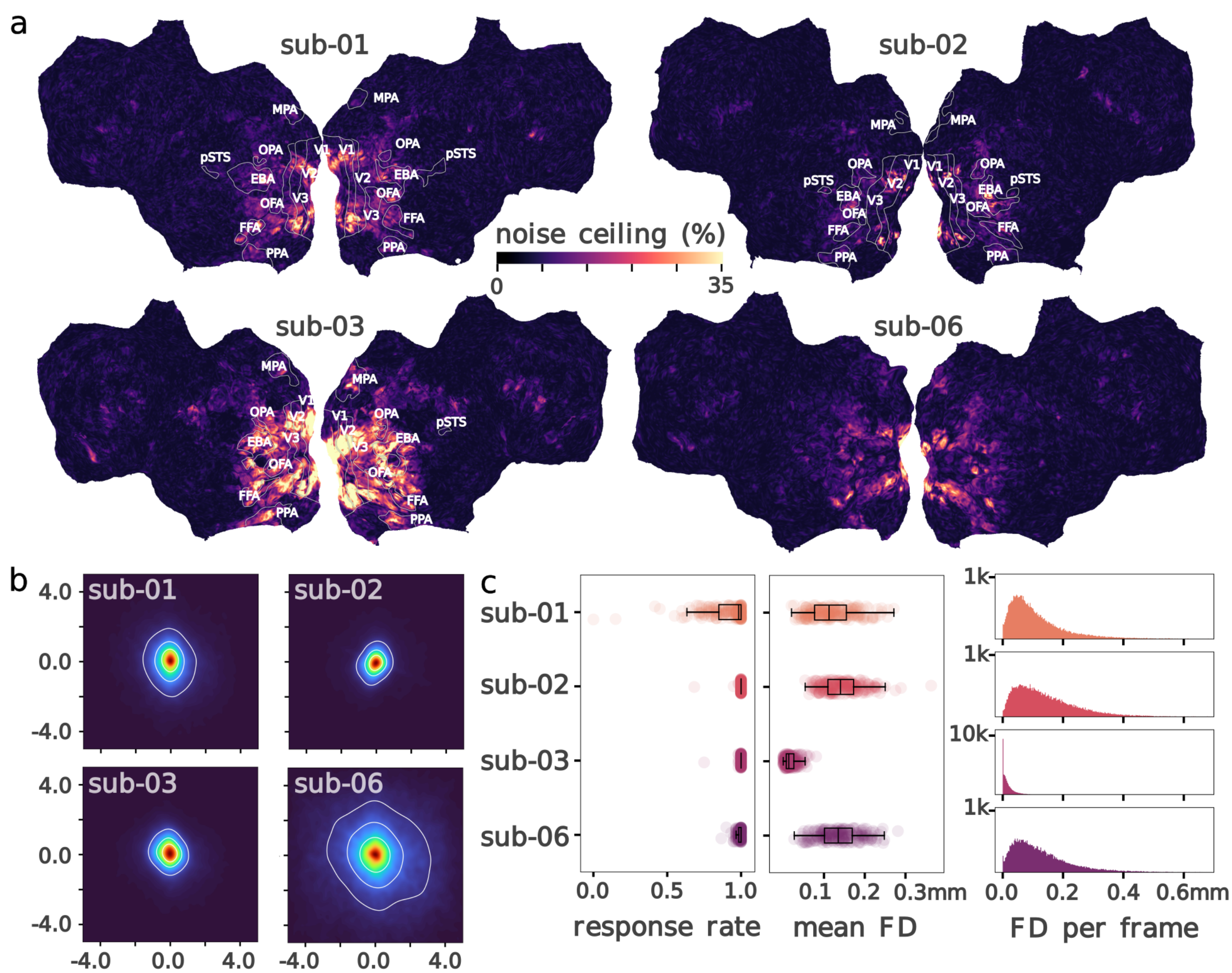

**Fig. 2. Quality metrics. a.** Voxelwise noise ceiling (% of variance explained) per participant shown on flattened cortical surfaces. For subjects who completed visual localizer tasks, the labeled outlines indicate functionally defined ROIs identified with the retinotopy (V1, V2 and V3) and fLoc (face preference: FFA, OFA and pSTS; body preference: EBA; scene preference: PPA, OPA and MPA) tasks. **b.** Gaze position in relation to the screen center, in ° of visual angle, during image presentation (data downsampled to 50Hz). Contours represent 25, 50 and 75% of the gaze density. **c.** Proportion of trials with a recorded behavioral response ("response rate") shown per run for each subject (left), and distribution of framewise displacement (FD, in mm) as an indication of head motion (middle: FD averaged per run; right: FD per frame).

### 2.10 Population Receptive Fields.

**Task design.** In addition to the main THINGS image-recognition task, three participants also completed multiple sessions (6 sessions for sub-01 and sub-02; 5 sessions for sub-03) of a retinotopy task adapted from Kay and colleagues (20) and implemented in Psychopy (23). These data were used to derive population receptive field (pFR) properties at the voxel level and to delineate ROIs from the early visual cortex. Each session included three functional runs of 301s (202 volumes at TR = 1.49s), each of which used a different aperture shape to stimulate the visual field: ring, bar or wedge. Each run included eight cycles of 31.2s during which an aperture moved slowly across the visual field to reveal a portion of visual pattern. Patterns were made of color objects shown at multiple spatial scales on a pink-noise background to drive both low-level and high-level visual areas. Patterns were drawn randomly at a rate of 15 frames per second from 100 different RGB images of $768 \times 768$ pixels (the Human Connectome Project retinotopy stimuli (35, 36)).

**Visual stimulation.** The stimulated visual field was a circular area whose diameter corresponded to 10 ° of visual angle in the center of the screen; i.e., $1280 \times 1024$ pixels covering $17.5 \times 14$ ° of visual angle. The exact procession of the pattern depended on its aperture.

- For ring runs, a thick circle aperture expanded from the center of the stimulated visual field for four consecutive cycles (each 28s of stimulation + 3.2s of rest), followed by a 12s pause, and then by four more cycles during which the ring contracted from the periphery to the center.
- For wedge runs, a rotating wedge aperture that corresponded to $1/4$ of the stimulated visual field completed four consecutive counter-clockwise rotations (each a 31.2s cycle), followed by a 12s pause and then by four more clockwise rotations.
- For bar runs, a wide bar aperture swept eight times (each 28s of sweep + 3.2s of rest) across the stimulated visual field, first from left to right, then from bottom to top, then right to left and top to bottom. After a 12s pause, the bar then swept diagonally from bottom left to top right, bottom right to top left, top right to bottom left, and top left to bottom right.

Participants were instructed to fixate their gaze throughout on a dot (diameter = 0.15 ° visual angle) presented centrally that alternated in color between blue and orange, and to press a button with the right thumb whenever the dot changed color using a custom MRI compatible video game controller (25).

**Population receptive field estimation.** Voxel-wise population receptive fields were estimated with the analyzePRF matlab toolbox (20) (commit a3ac908 based on release 1.6) in matlab R2021a. We used temporal averaging to downsample binary aperture masks and obtain TR-locked (TR = 1.49s) binary masks which we resized to $192 \times 192$ pixels to reduce processing time. For each subject, BOLD data were preprocessed with the fMRIPrep pipeline (28, 29) (version 20.2.6), detrended, normalized and averaged over repeated runs of the same type (e.g., ring aperture). The first three volumes of each run were dropped to allow for signal equilibrium. Whole-brain voxels were vectorized and split into chunks of up to 240 voxels each, and processed in parallel with analyzePRF, after which voxelwise output metrics were reassembled into volumes. Receptive field sizes and eccentricities were converted from pixels into ° of visual angle, while angles were converted from compass to signed north-south. Volumes were converted into surfaces with FreeSurfer mri_vol2surf, and visual ROI boundaries (V1, V2, V3, hV4, VO1/VO2, LO1/LO2,TO1/TO2, and V3a/V3b) were estimated with Neuropythy (37) (version 0.11.9) using a bayesian mapping approach that refines individual parameters with group atlas priors. Surface values were reconverted into volumes in functional native subject space (T1w) with FreeSurfer's mri_convert, FSL's fslreorient2std and Nilearn (38)'s resample_to_img. The current data release includes binary ROI masks in native (T1w) volume space for sub-01, sub-02 and sub-03.

### 2.11 Functional localizer (fLoc).

**Task design.** Three subjects (sub-01, sub-02 and sub-03) also completed six sessions of a functional localizer task to identify brain regions responding preferentially to specific stimulus categories. The task was based on a Psychopy implementation (https://github.com/NBCLab/pyfLoc) of the Stanford VPN lab's fLoc task (19) using stimuli from the fLoc functional localizer package downloaded from https://github.com/VPNL/fLoc (commit 9f29cbe). Each session included two functional runs of 231s (155 volumes at TR = 1.49s) with randomly ordered 5.96s blocks of rapidly presented images from one of five categories : faces, places, body parts, objects and characters. Each block included 12 trials for which a $768 \times 768$ image from the block's category was displayed centrally for 0.4s, followed by a 0.095-0.1s ISI. Subjects were instructed to fixate on a red dot shown in the middle of the screen throughout the run. To maintain engagement, they were instructed to press a button on a custom MRI compatible video game controller (25) with the right thumb whenever the same image appeared twice in a row; i.e., the "one-back" task variation. Blocks during which no image was shown—only the red fixation dot appeared on a grey background for 5.96s—were intermixed in the block sequence to estimate a baseline condition. Each functional run included 6 blocks from each of the five image categories and 6 blocks of baseline. The first run of each session featured images from the house (places), body (body parts), word (characters), adult (faces) and car (objects) sub-categories from the fLoc package; the second run featured images from the corridor (places), limb (body parts), word (characters), adult (faces) and instrument (objects) sub-categories.

**ROI delineation.** Functional data were preprocessed with fMRIPrep (version 20.2.5) (28, 29) and analyzed in native space (T1w) with a general linear model implemented in nilearn 0.9.2 (38). Each run's first three functional volumes were dropped for signal equilibrium. Data were fitted with the canonical SPM HRF using a cosine drift model and an autoregressive noise model of order 1. Regressed out confounds included the mean global, white matter and CSF signal as well as the six basic head motion parameters. Data were normalized (z-scored within voxel along the time dimension), spatially smoothed (FWHM = 5mm) and masked with a binary mask created from the intersection of all 12 run-specific functional masks outputted by fMRIPrep. The following t-contrasts were applied to identify category-specific ROIs:

- **face > (bodies + characters + places + objects) :**

fusiform face area (FFA), occipital face area (OFA) and posterior superior temporal sulcus (pSTS)

- **place > (face + bodies + characters + objects) :** parahippocampal place area (PPA), occipital place area (OPA) and medial place area (MPA)

- **bodies > (face + character + place + object) :** extrastriate body area (EBA)

To delineate ROI boundaries (FFA, OFA, pSTS, EBA, PPA, OBA and MPA), clusters from subject-specific maps of fLoc t-contrasts were intersected with an existing group-derived parcellation of category-selective brain regions made accessible by the Kanwisher lab (39, 40). Binary group parcels were warped from CVS (cvs_avg35) to MNI space with Freesurfer 7.1.1 and from MNI to T1w space with ANTs 2.3.5. Each subject-specific ROI was identified within a mask of thresholded clusters from the relevant t-contrast map (e.g., face > other conditions for FFA; alpha=0.0001, t > 3.72, clusters > 20 voxels) that intersected with the smoothed group parcel (mask values > 0.01 post spatial smoothing, fwhm=5 mm). Within this intersection mask, voxels with the highest t-values (at least 3.72) were selected in proportion to the warped group parcel size (up to 80% of the pre-smoothing voxel count). The current data release includes binary ROI masks in native volume space (T1w) for sub-01, sub-02 and sub-03.

### 2.12 ROIs on cortical flat maps.

For visualization purposes, ROI boundaries delineated with fLoc for the FFA, OFA, pSTS, EBA, PPA, OBA and MPA, and ROI boundaries estimated with Neuropythy for V1, V2 and V3 were projected onto flat cortical surfaces using Pycortex 1.2.5 (32), and drawn manually in Inkscape 1.3.2. For V1, V2 and V3, voxels with estimated eccentricities greater than 10 ° of visual angle were masked out, restricting the final ROIs to reflect the portion of visual field stimulated by our retinotopy paradigm. Individual ROI boundaries can be made visible as annotations on the cortical flat maps that are parts of this data release to help interpret the location of brain activity patterns (e.g., Fig. 2a).

## 3. Data Records

We use DataLad (41), a data version control tool built on top of git and git-annex, to track the provenance of all data assets in this release. The data, documentation and code are organized as a nested set of DataLad submodules inside the https://github.com/courtois-neuromod/cneuromod-things repository. The repository structure and content are detailed in the main README.md file.

The data included in this collection can be downloaded with DataLad without requiring registered access (see section 5). The four participants have requested access to their own data, and have released them openly as citizen scientists under a liberal Creative Commons (CC0) data license via the data portal of the Canadian Open Neuroscience Platform (42) (CONP). Non-identifying raw and derivative data are available, while identifying files (i.e., detailing scan dates) are only shared in their anonymized form. BOLD data are organized in the Brain Imaging Data Structure (BIDS) standard (43).

Files related to the main THINGS image recognition task are found under cneuromod-things/THINGS:

- `cneuromod-things/THINGS/fmriprep` includes raw and preprocessed BOLD data, eye-tracking data, *events.tsv files with trial-wise metrics (stimulus- and task-related), image stimuli, and stimulus annotations.

- `cneuromod-things/THINGS/behaviour` includes analyses of the subjects' fixation compliance and performance on the continuous recognition task.

- `cneuromod-things/THINGS/glmsingle` includes fMRI analyses and derivatives, including trial-wise and image-wise beta scores estimated with GLMsingle (33), voxel-wise noise ceilings, and data-driven analyses to showcase the quality of the data.

- `cneuromod-things/THINGS/glm-memory` includes GLM-based analyses of memory effects in the preprocessed BOLD data and associated statistical maps.

In addition, the CNeuroMod-THINGS dataset includes data, scripts and derivatives from the two vision localizer tasks completed by three of the four subjects (sub-01, sub-02 and sub-03), which we used to derive subject-specific ROIs. Those files are found under `cneuromod-things/fLoc` and `cneuromod-things/retino` for the functional localizer and retinotopy tasks, respectively. The `cneuromod-things/anatomical` submodule also includes patch files and instructions to project voxel-wise statistics from any CNeuroMod dataset onto subject-specific flattened cortical surfaces (flat maps) for visualization. Surfaces feature manually traced outlines of visual ROIs for subjects who completed the fLoc and retinotopy localizers.

Finally, `cneuromod-things/datapaper` includes jupyter notebooks to recreate figures from the current manuscript using source data and result files from the relevant DataLad submodules.

## 4. Technical Validation

### 4.1 Data quality metrics.

**Response rate.** Response rate was high across participants, ranging from 91.13% (sub-01) to 99.84% (sub-03). For each subject, a majority of runs had near perfect response rates (median run response rate > 98%), although the number of runs with lower response rates was higher for sub-01 than for the other participants due to self-reported bouts of drowsiness (Fig. 2b). The number of trials with no recorded response out of 12,780 trials was 1134 (sub-01), 24 (sub-02) and 21 (sub-03), and 96 out of 11,700 trials (sub-06).

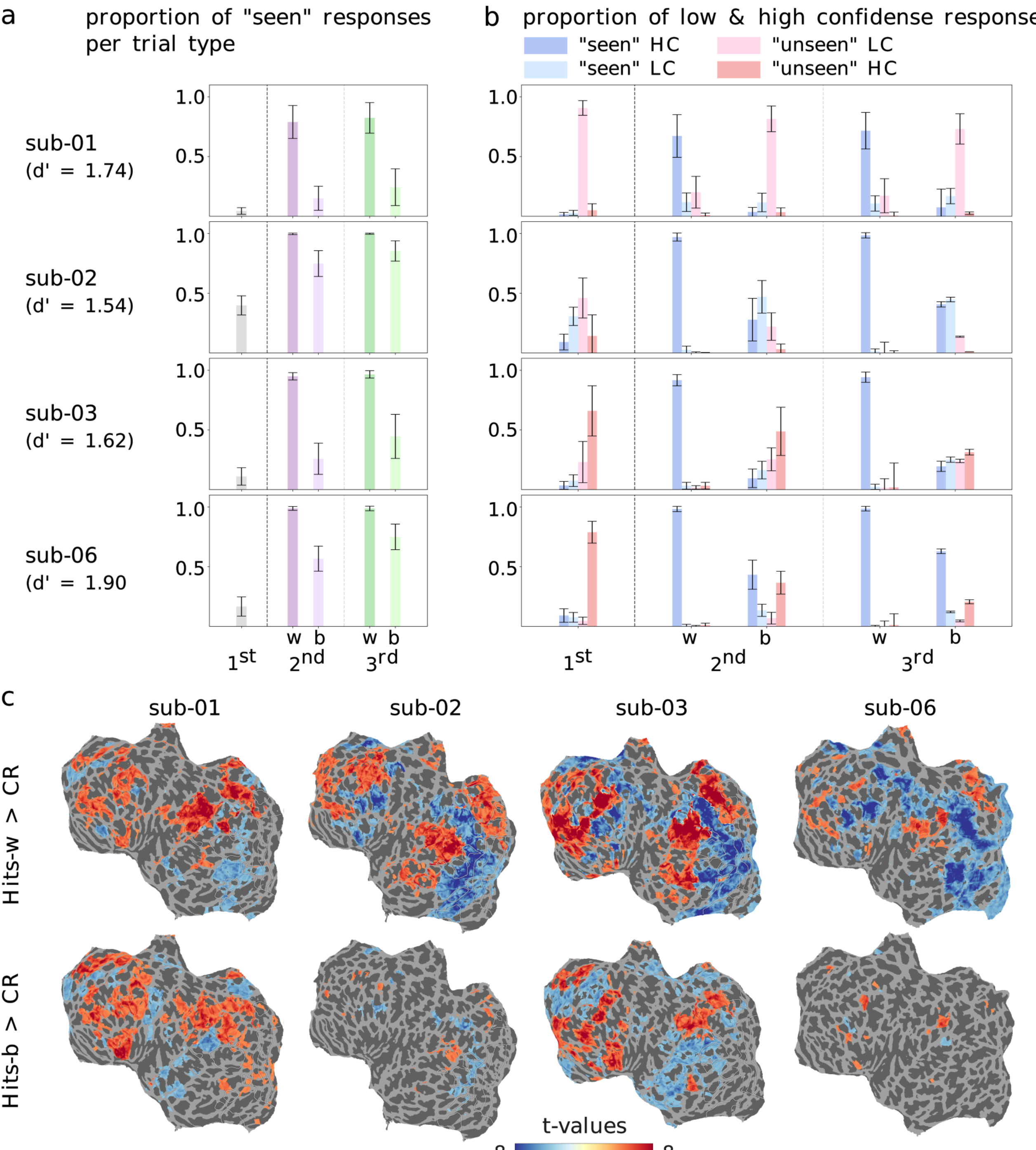

**Fig. 3. Memory Recognition. a.** Proportion of "seen" answers per image repetition, averaged across sessions for each subject. For the 1st image presentation (left, in grey), "seen" answers are false alarms. For the 2nd and 3rd image presentations (purple and green, respectively), "seen" answers are hits, and response rates are split between images repeated within (w, darker shades) and between (b, paler shades) sessions. Error bars are standard deviations. **b.** Proportion of answer types per image repetition (1st, 2nd and 3rd presentation), averaged across sessions for each subject. Responses include "seen" and "unseen" answers split between low and high confidence (LC and HC). Error bars are standard deviations. For the 2nd and 3rd presentation, results are split between images repeated within (w) and between (b) sessions. "Seen" answers (high confidence in darker blue, low confidence in pale blue) are incorrect—false alarms—for the 1st rep, and correct—hits—for the 2nd and 3rd reps. "Unseen" answers (high confidence in red, low confidence in pink) are correct—correct rejections—for the 1st rep, and incorrect—misses—for the 2nd and 3rd reps. **c.** Thresholded t-scores per participant for two-sample t-tests contrasting trial-wise beta scores between memory conditions shown on flattened cortical surfaces (betas concatenated across all sessions, $p < .0001$ uncorrected, unequal variance assumed). Top panel results compare BOLD responses for "within-session hits" (positive values) and for "correct rejections" (negative values). Bottom panel results compare BOLD responses for "between-session hits" (positive values) and for "correct rejections" (negative values).

**fMRI data.** We assessed the intrinsic quality of the fMRI data using framewise displacement (FD, in mm) as a measure of head motion. FD distributions indicate low levels of motion in each participant. We report a majority of frames with less than 0.1mm FD, and a mean FD per run inferior to 0.15mm for a majority of runs in all subjects (Fig. 2d). We further note that sub-03 demonstrated exceptionally low levels of motion with a mean FD below 0.1mm. To assess task-evoked signal, we calculated noise ceilings (section 2.9) as an estimate of the percentage of signal explained by stimulus images in each voxel (Fig. 2a). Higher noise ceilings were observed in low-level visual areas (V1, V2 and V3, identified via pRF; section 2.10) as well as visual cortical regions with known categorical preferences like the FFA, EBA and PPA (identified with fLoc; section 2.11), indicating consistent stimulus-specific signal in the BOLD data. Maximal noise ceilings were 56.51 (sub-01), 66.43 (sub-02), 73.89 (sub-03), and 54.19 (sub-04). We note that the exceptionally low levels of motion in sub-03 likely contributed to their high noise ceilings.

**Visual fixation.** We used eye-tracking data to estimate fixation compliance during image viewing (section 2.6). We performed quality checks to exclude runs with missing, corrupt or low quality (i.e., very noisy) data. The proportion of runs with usable, drift-corrected eye-tracking data was generally high for each subject: 140/213 (sub-01), 188/213 (sub-02), 190/213 (sub-03), 164/195 (sub-06). The distribution of drift-corrected gaze position in relation to the fixation marker during image viewing indicates high levels of fixation compliance in sub-01, sub-02 and sub-03, and sub-06 to a lesser extent (Fig. 2b).

### 4.2 Memory effects.

**Behavioural performance.** To determine whether subjects recognized images above chance level, we computed d', the standardized difference between the hit rate—the proportion of previously shown items correctly recognized as "seen"—and the false alarm rate—the proportion of items shown for the first time wrongly identified as "seen". High d' scores indicate that all subjects performed above chance : 1.744 (sub-01), 1.536 (sub-02), 1.623 (sub-03), 1.898 (sub-06). Predictably, hit rates show that more images were correctly recognized when repeated within rather than between sessions (Fig. 3a), indicating greater task difficulty at longer delays (days versus minutes). This effect was dampened in sub-02 whose hit rate was closest to ceiling, although this subject's responses also included the highest number of false alarms. Faster response times were also observed for within-session hits compared to between-session hits for all subjects (Fig. S1).

The distribution of response types ("seen" and "unseen" answers given with low and high confidence) per condition further illustrates the difficulty of recognizing images after longer delays. For all subjects, within-session repetitions were mostly high confidence hits (previously seen images correctly labeled as "seen"; Fig. 3b). Between-session repetition trials included greater proportions of low confidence hits, and of low or high confidence misses (previously seen images incorrectly labeled as "unseen"), indicating much weaker memory. The impact of repetition delays was observed for both second and third image presentations. In fact, the distribution of response types for images repeated between sessions (labelled "b" under 2nd and 3rd reps, Fig. 3b) is comparable to the response profile of first-time image presentations (1st rep) for which there is no memory (identical response distributions between seen and unseen conditions indicate chance level). Of note, the response profile of sub-06 indicates the strongest memory signal for between-session repetitions, as it is most distinct from the response distribution for first image presentations.

**fMRI signal.** Contrasting BOLD activity patterns between memory conditions also highlights more salient memory effects for images repeated within- rather than between-sessions. We performed two-sample t-tests contrasting trial-wise beta scores (estimated with GLMsingle and concatenated across sessions) associated with hits (correctly recognized images) and correct rejections (never seen images correctly identified as "unseen"). The results (Fig. 3c) reveal widespread deactivation in visual cortical areas for within-session hits compared to first-time presentations. This "repetition suppression effect" could be mediated by neural fatigue at very short delays (e.g., for consecutive trials), and by familiarity, attention, perceptual expectations and response time at slightly longer delays (44–46). Of note, this effect was greatly reduced when contrasting between-session hits to first-time presentations. Successful memory recognition was also associated with enhanced prefrontal and parietal activation.

To assess within-run memory effects, we also performed fixed-effects analyses on first-level GLMs applied to fMRIPrep preprocessed run-level BOLD data using Nilearn (38). These analyses, whose resulting t-values are included in the current data release, revealed non-significant patterns similar to those shown in Fig. 3c when contrasting within-session hits and between-session hits to correct rejections. This lack of significance indicates high variability in memory effects across runs and sessions. Further modelling of memory effects that takes different sources of variability into account is therefore warranted.

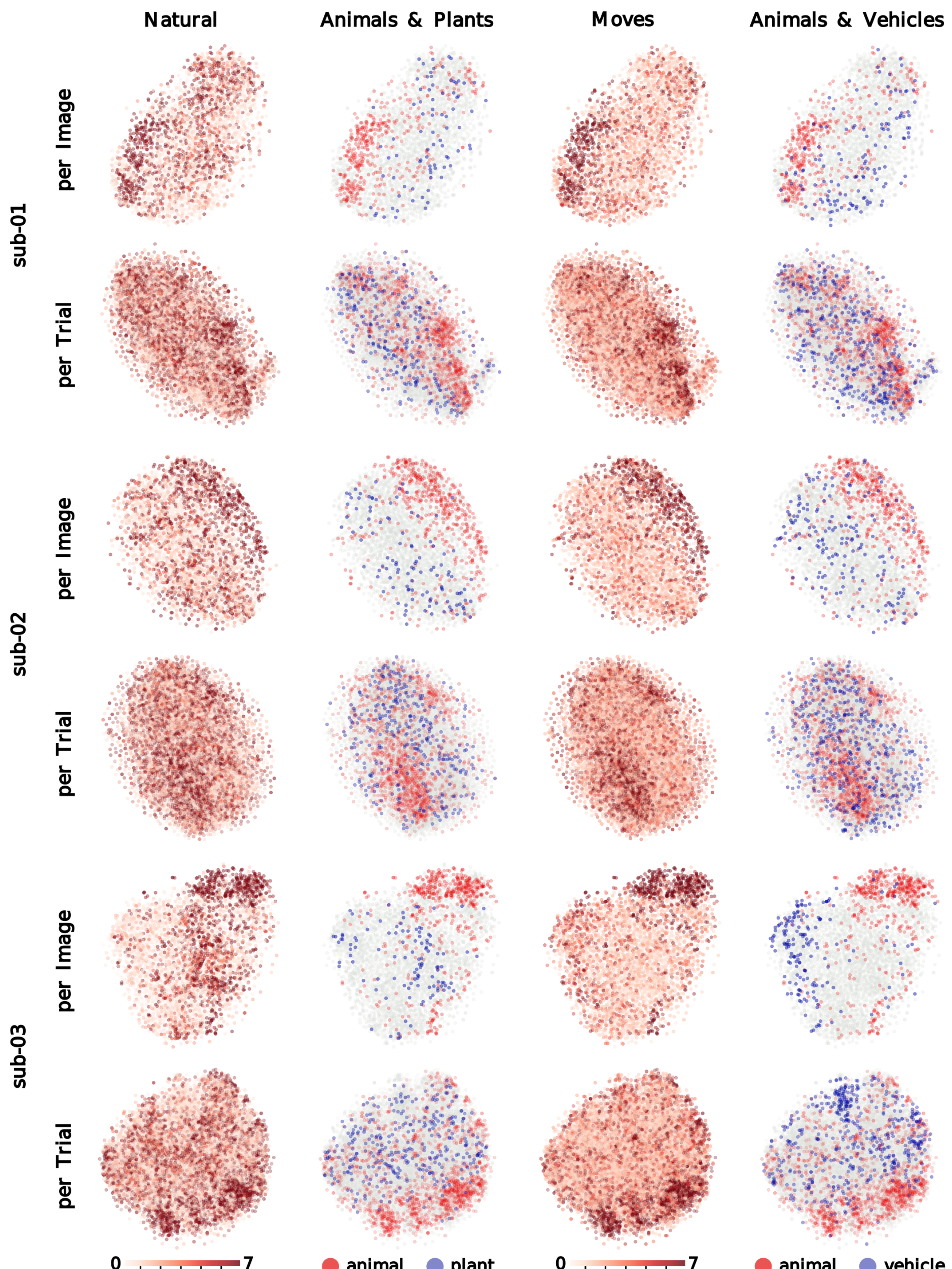

**Fig. 4. Dimension reduction analyses (t-SNE).** T-SNE plots on 50 principal components derived from normalized beta scores per trial and per image from functionally-defined, subject-specific ROIs with category-specific voxel signal (t > 2.5 for either the face, the body or the scene fLoc contrast from unsmoothed BOLD data). ROIs included 1345, 1224 and 1817 voxels for sub-01, sub-02 and sub-03, respectively. Image content is annotated with categorical labels (i.e., "animal", "plant", "vehicle") and object concept ratings ranging between 0 and 7 (i.e., "moves", "natural") from the THINGSplus project (9).

### 4.3 Dimensionality reduction analyses (t-SNE).

We conducted data-driven dimensionality reduction analyses to visualize the representation of semantic information in brain regions with categorical preferences. Specifically, we generated t-distributed stochastic neighbor embedding (t-SNE) plots (47) from beta scores estimated with GLMsingle (section 2.7) within subject-specific ROIs. ROI boundaries were delineated using the following criteria: all voxels with t > 2.5 for either the face, the body or the scene fLoc contrast from unsmoothed BOLD data. ROI extents included 1345 voxels (sub-01), 1224 voxels (sub-02), and 1817 voxels (sub-03). For each subject, one t-SNE plot was generated from trial-specific beta scores (excluding no-response trials) and another from image-specific beta scores averaged across repetitions (including only images with three repetitions with recorded responses). Beta scores were z-scored within voxel and then reduced with PCA, keeping the top 50 components (PCs). T-SNE plots generated with Scikit-learn 1.0.1 were initialized with the betas' first two PCs scaled by 0.0001 of the first PC's standard deviation (perplexity = 100, learning rate = 500, max 2000 iterations).

We used higher-order WordNet categorical labels ("animal", "plant", "vehicle") and object concept ratings between 0 and 7 (i.e., "moves", "size", "natural") from the THINGSplus project (9) to annotate image content within the plot clusters. Fig. 4 showcases annotated t-SNE plots derived from image-wise (top row) and trial-wise (bottom row) beta values for sub-01, sub-02 and sub-03. Clustering patterns indicate greater coherence for image-wise signal, suggesting that averaging over repetitions reduces noise in stimulus-specific signal and increases the robustness of analyses conducted at the item level. Additional t-SNE plots generated with beta values from face-specific ROIs (FFA and OFA), from place-specific ROIs (PPA, OPA, and MPA), and from the early visual cortex (V1, V2 and V3) are included in Supplementary S5—Dimensionality reduction analyses in functionally defined ROIs.

### 4.4 Beta distributions within single ROI voxels.

We assessed whether stimulus image content—i.e., the presence of faces and the complexity of scene elements in the image—influenced the distribution of image-specific beta scores (section 2.7) within functionally defined ROIs (section 2.11) known for the preference for faces (FFA), body parts (EBA) and scenes (PPA). For sub-06—who did not complete the fLoc task—ROIs were estimated using group-derived binary parcels of the FFA, EBA, and PPA (39, 40) warped to the subject's T1w space with Freesurfer 7.1.1 and smoothed with a FWHM=3 kernel.

Within each ROI, we first identified the voxel with the highest noise ceiling (section 2.9). We then split the voxel's image-specific beta scores into separate distributions based on the type of face contained in the image (human or animal; Fig. 5, in blue), and then based on whether the image depicted a full-blown scene, an object in a rich or in a minimalist background or a lone object (Fig. 5, in green). The split was based on boolean image annotations generated manually by author MSL (Supplementary Table S1). Only images with three repetitions with recorded button presses were included in this analysis. Beta distributions reveal a clear preference for images that contain faces (human or otherwise), and an indifference to scene elements, in both the FFA and the EBA voxel across subjects—keeping in mind that faces and body parts frequently co-occur in natural images. Meanwhile, beta distributions from the PPA voxel indicate a preference for complex scene elements, and a slight preference for the absence of faces. These results illustrate how clear categorical preferences can be observed for single images at the voxel level in unsmoothed BOLD signal in the current dataset.

## 5. Usage Notes

The CNeuroMod-THINGS dataset is available as a DataLad collection on GitHub :
https://github.com/courtois-neuromod/cneuromod-things.

The four subjects have requested access to their data and chosen to share them openly via the data portal of the Canadian Open Neuroscience Platform (42) (CONP; https://portal.conp.ca/dataset?id=projects/cneuromod) as citizen scientists. The data are distributed under a liberal Creative Commons (CC0) data license that authorizes the resharing of derivatives. You can download the data from the CONP portal without registered access.

You will need the DataLad (41) software (version > 1.0.0, https://www.datalad.org/), a tool for versioning and accessing large data structures organized in git repositories available for Linux, OSX and Windows. For secure data transfers, we recommend using SSH protocols by creating an SSH key on the machine where the dataset will be installed, and adding the key to your GitHub account.

First, clone the dataset repository onto your local machine:

```
datalad clone git@github.com:courtois-neuromod/
cneuromod-things.git
```

A warning message will be thrown because the remote origin does not have git-annex installed. This issue will not prevent the installation. You can now download the repository data. The cneuromod-things repository is a nested collection of git submodules whose overall structure is detailed in the main README.md file. When you first clone the CNeuroMod-THINGS repository, submodules will appear empty (e.g., cneuromod-things/THINGS/glmsingle). To download a specific data subset, you can navigate to the submodule whose content you need and pull the files directly from there. Use the `datalad get` command once to download the submodule's symbolic links and files stored directly on GitHub, and then a second time to pull files from the remote CONP server.

For example, the commands below download

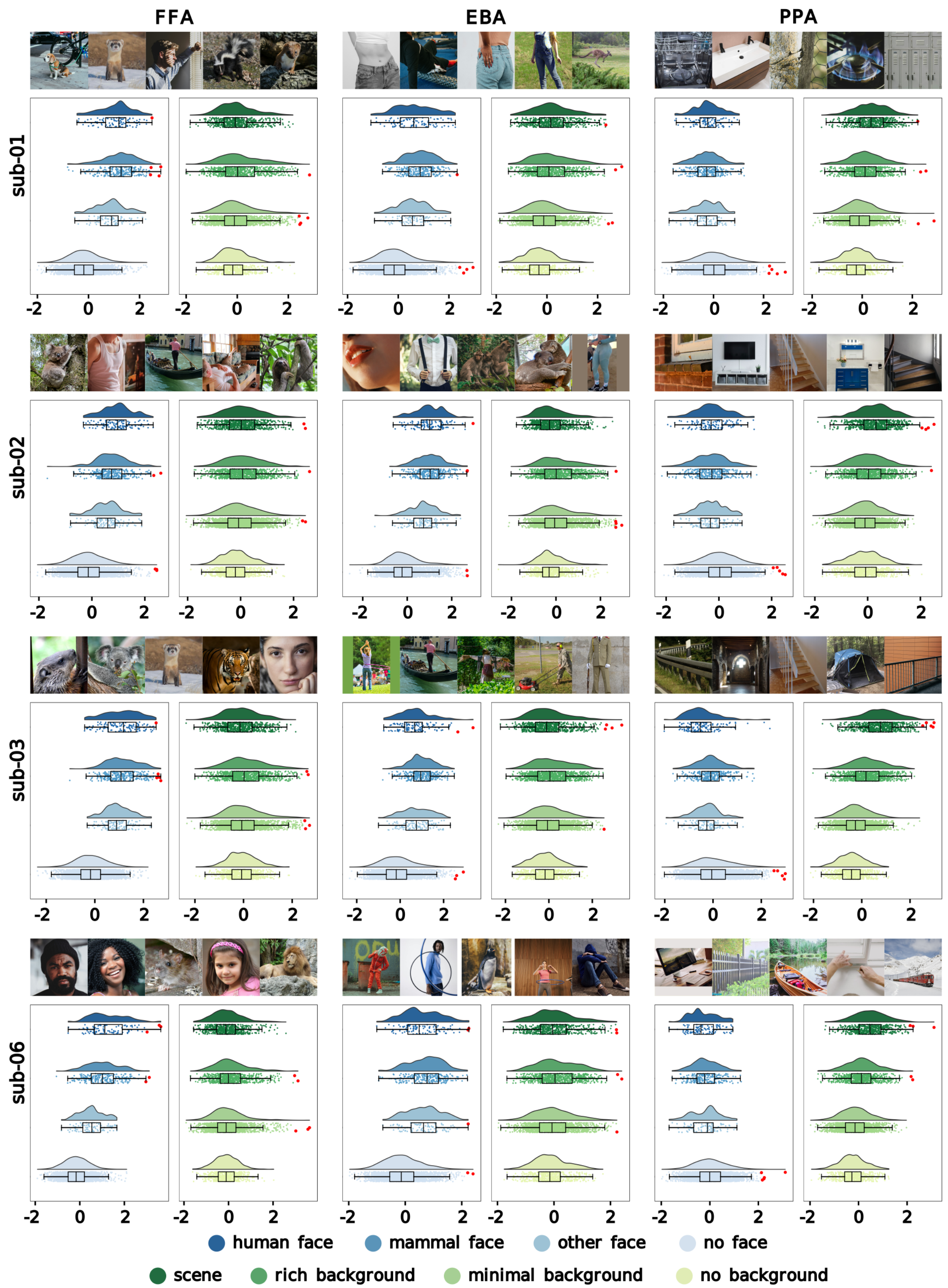

**Fig. 5. Voxel-level image-specific beta distributions within functionally defined ROIs.** ROIs include the FFA (left), EBA (middle) and PPA (right). For each subject, charts show image-specific betas from the voxel with the highest noise ceiling in the ROI split into separate distributions based on image content. In blue, betas are split according to the presence of human, non-human mammal, and other faces (e.g., bird or insect) within an image's central focus. In green, betas are split according to whether the image is a scene, has a rich background, features some central object(s) with minimal background, or is a lone object without any background. For each subject and ROI, the images with the five highest beta scores are shown as red dots in the distributions. Five public domain or CC0 images that resemble these dataset images are shown above each set of distributions. See Supplementary section S3 for the list of THINGS stimulus images with the twelve highest beta values for each ROI voxel.

files saved under `sub-01/qc` in the cneuromod-things/THINGS/glmsingle submodule.

```
cd cneuromod-things/THINGS/glmsingle
datalad get *
datalad get -r sub-01/qc/*
```

While it is technically feasible to pull the entire content of all nested submodules recursively with a single command (with `datalad get -r cneuromod-things/*`), we strongly recommend against it due to the complexity and depth of the nested repository structure and sheer dataset size.

See our official documentation for additional information on accessing CNeuroMod datasets.

## 6. Data Availability

The data, documentation and code can be accessed from https://github.com/courtois-neuromod/cneuromod-things, a repository of nested submodules, using DataLad without registered access. Data are released under a liberal Creative Commons (CC0) license that authorizes the re-sharing of derivatives.

## 7. Code Availability

BOLD data were acquired with the PsychoPy library and preprocessed with the fMRIprep pipeline (28, 29) (versions 20.2.3 and 20.2.5 ; https://fmriprep.org/en/stable/). All CNeuroMod data acquisition and data preprocessing scripts are available on the CNeuroMod GitHub (https://github.com/courtois-neuromod):

- Data acquisition scripts :
  https://github.com/courtois-neuromod/task_stimuli
- BOLD data preprocessing scripts :
  https://github.com/courtois-neuromod/ds_prep

The code used to generate derivatives from the preprocessed CNeuroMod-THINGS data is integrated in the https://github.com/courtois-neuromod/cneuromod-things repository and its submodules. This repository includes scripts to:

- extract trial-wise and image-wise beta scores per voxels from preprocessed BOLD data
  `THINGS/glmsingle/code/glmsingle`
- perform t-tests and GLM fixed-effects analyses to assess memory recognition effects on the BOLD data
  `THINGS/glm-memory/code`
- quantify in-scan head motion
  `THINGS/glmsingle/code/qc`
- organize trial-wise metrics (stimulus image annotations, task conditions, task accuracy, reaction time, gaze fixation compliance)
  `THINGS/fmriprep/sourcedata/things/code`
- analyze behavioural performance on the image recognition task
  `THINGS/behaviour/code`
- process eye-tracking data
  `THINGS/fmriprep/sourcedata/things/code`
- perform data-driven analyses to characterize stimulus representation in the brain signal
  `THINGS/glmsingle/code/descriptive`
- derive ROI masks from two functional localizer tasks
  `retinotopy/prf/code` & `fLoc/rois/code`
- visualize results onto flattened cortical maps of the subjects' brains
  `anatomical/pycortex`

The CNeuroMod-THINGS repository also provides Jupyter Notebooks (`datapaper/notebooks`) to reproduce the figures included in the current manuscript directly from the source data pulled from the DataLad collection.

#### AUTHOR CONTRIBUTIONS

**MSL:** Methodology, Software, Data curation, Formal analysis, Visualization, Writing – original draft, review & editing
**BP:** Conceptualization, Project administration, Investigation, Methodology, Software, Data curation, Formal analysis, Visualization
**OC:** Software, Formal analysis, Visualization, Writing – review & editing
**ED:** Software, Formal analysis, Data curation, Writing – review & editing
**KS:** Software, Writing – review & editing
**VB:** Conceptualization
**JB:** Conceptualization, Funding acquisition, Resources, Project administration, Investigation, Data curation, Writing – review & editing
**LB:** Conceptualization, Funding acquisition, Resources, Project administration, Investigation, Data curation, Formal analysis, Visualization, Writing – review & editing
**MNH:** Conceptualization, Funding acquisition, Supervision, Methodology, Software, Data curation, Formal analysis, Visualization, Writing – original draft, review & editing

#### COMPETING FINANCIAL INTERESTS

The authors declare no conflict of interest.

#### ACKNOWLEDGEMENTS

This work was supported by the Courtois Foundation and an NSERC discovery grant awarded to LB (RGPIN-2025-06022), and by a Max Planck Research Group Grant (M.TN.A.NEPF0009) and an ERC Starting Grant COREDIM (StG-2021-101039712) awarded to MNH.

# Supplementary Material

## S1. Misordered sessions

Due to errors at the console, a small number of sessions were administered out of their pre-planned order, which introduced irregular patterns of repetition (in delays and rep numbers) for some images. Data from affected sessions may include unexpected memory interactions; as such, we recommend that they be omitted from memory-specific analyses. Irregular sessions are flagged to facilitate this decision. The "atypical", "atypical_log" and "not_for_memory" columns in run *events.tsv files can be used to filter out trials (columns are defined in this document: task-things_events.json)

Specific session-level deviations from the pre-planned presentation order are described below.

- **sub-01, sess-17.** This session's planned runs were accidentally administered in the following order: 1, 2, 3, 5, 6, 4 (as labeled in the dataset). The condition ("seen"/"unseen") and sub-condition (e.g., "seen-between", "seen-between-within") and the subject's performance metrics ("error") are corrected in the *events.tsv files to reflect the order in which stimuli were shown (rather than planned). This mistake has minimal impact on the overall structure of the experiment, and no impact on previous or subsequent sessions. Sub-01, ses-17 also included a "false-start" for which a run was interrupted and restarted, introducing additional repeats for that segment of the session. Condition labels and performance metrics were corrected where appropriate for the few trials affected by these additional repeats (e.g., no error was counted if the subject recognized an image shown during the false-start).
- **sub-03, sess-14.** Stimuli prepared for sub-02, ses-04 were accidentally shown instead of the planned sub-03, sess-14; data from this accidental presentation were dropped. Sub-03, ses-14 was correctly re-run two weeks later with the proper stimuli, as included in the dataset. This mistake introduced a larger delay between sub-03 ses-13 and ses-14 and potential interference in the memory task by showing additional stimuli between sessions meant to be consecutive. However, it did not alter repetition patterns for any of the previous or subsequent sessions.
- **sub-03, ses-22 to ses-25.** The planned sub-03, ses-25 was accidentally run instead of sub-03 ses-22. After this mistake, ses-22 (with the correct stimuli), ses-23, ses-24, and ses-25 (repeated a second time in its proper order) were administered with their planned stimuli to correct course. This mistake introduced delays and interference between ses-21 and ses-22, and additional image repeats that required correcting condition labels and performance metrics for ses-24, ses-25 and ses-26 (e.g., introducing higher numbers of repetitions for some images, and atypical sub-condition labels like "seen-within-between-within-between"). Data from the first ses-25 (administered out-of-order) are dropped from the dataset.
- **sub-06, ses-20.** Stimuli from sub-01 ses-20 were accidentally shown for sub-06 session 20. After this mistake, sub-06 ses-21, ses-22 and so-on were then run with their planned stimuli. Data from ses-20 administered with the wrong stimuli were kept in the dataset (no repeat was made of ses-20 with the planned stimuli). By chance, the stimuli accidentally shown in sub-06 ses-20 were planned for that 6-session block so that task deviations were contained to that block. Thus, only ses-20 to ses-26 required corrections; e.g., introducing higher numbers of repetitions and atypical sub-condition labels. All sessions (20-26) are included in the dataset with corrected labels and performance metrics.
- **sub-06, ses-19.** Due to scanner issues, sub-06 ses-19 included two "false-starts" for which a run was interrupted and restarted. This issue introduced additional repeats for the task segments that were shown twice. Corrections were made to conditions and performance metrics for the few trials affected by these additional repeats.

## S2. Manual Image annotation

Table S1 defines the boolean flags introduced by author MSL to annotate the content of the stimulus set.

| Flag name | Description |
|---|---|
| face | Contains any face, whole or partial (e.g., eyes or smile), central or in the periphery (incidental to the image's main focus). Faces can be human or not, real or artificial (a doll's face, a cartoon cat face, a face reflected in a soap bubble). |
| body | Contains any body (which may include a face) or non-face body part(s), central or in the periphery. Bodies can be human or not, real or artificial (a robot's body, an octopus tentacle, an artificial limb), clothed or not (e.g., a person's legs in an image featuring pants, a gloved hand holding a wine glass, a head with visible shoulders and torso). |
| human face | Contains any face or portion of a face with human features, central or in the periphery, real or artificial (a doll's face, a cartoon of a person's face, etc). |
| human body | Contains any human body or non-face human body part, central or in the periphery, real or artificial (a drawn hand). |
| non-human mammal face | Contains any face with non-human mammalian features, central or in the periphery, real or artificial (a cartoon dog face). **Includes:** faces of rodents, dogs, felines, cows, deers, etc. **Excludes:** faces of insects, reptiles, fish, birds, sea mammals* (e.g., dolphins and whales). |
| non-human mammal body | Contains any body or non-face body part from a non-human mammal, central or in the periphery, real or artificial (a robot dog leg). **Includes:** rodents, felines, cows, deers, etc. **Excludes:** insects, reptiles, fish, birds, sea mammals* (e.g., dolphins and whales). |
| central face | Contains a face or a portion of a face (human or not, real or artificial) within the image's central focus. E.g., An image featuring a person jumping on a pogo stick is considered central if the face is visible, while faces of random spectators in the background are not. Although the face does not need to be in the middle of the image to be "central" per se, the face should be visible when gazing at a central fixation cross, and it must be part of the image's main focus of interest. |
| central body | Contains a body or non-face body part within the image's central focus. E.g., an image featuring a hand holding an item of interest is central, while a silhouette visible next to an aircraft carrier seen from afar is not. Although the body (part) does not need to be in the center of the image to be "central", it should be visible when gazing at a central fixation cross, and it must be part of the image's main focus of interest. |
| artificial face | Contains any representation of a human, animal or humanoid face that is not a real (living) face. Eg., a doll face or mannequin head, a cartoon bird face, a robot with facial features like eyes and a smile, an action figure, a painting or a photo of a face on a banner. |
| artificial body | Contains any representation of a human, animal or humanoid body or body part that is not a real (living) body (part). Eg., a statue, a prosthetic leg, a robotic hand, a drawing of a bird on a teapot. |
| scene | An item pictured in an environment with a background and a foreground that gives the sense of a place around it. A scene includes scenery, a view point and some perspective. It can be the image of a large object in a specific setting, like an aircraft in a hangar, a sofa in the middle of a living room, an anchor by a waterfront, an elephant at the zoo or a person skateboarding in a busy park. |

| Flag name | Description |
|---|---|
| rich background | An image of an object taken from closer than a scene, but that still includes items, people or animals clearly visible in the background. E.g., a backpack on someone's back walking away toward some trees, a tool in a garage with equipment visible behind it, a plant in a garden surrounded by other plants, a beer glass held by a person sitting between others, an apple on a table with orchard trees behind it. |
| lone object | The featured object is shown centrally with no additional objects visible in the periphery or background. Not only is the object shown by itself, but the empty background is uniform and minimally textured (no carpet, dinner mat or wooden fence behind) or blurred so that only the lone object is in focus. Note that objects can either be shown with zero background (the "lone objects" flag), with minimal background (e.g., an apple in a basket on a table), with noticeable background (the "rich background" flag) or within a scene (with background and perspective). |

**Table S1. Manual image annotation flags and their definitions.**
*Although they are mammals, sea mammal's features are very different from those of humans, hence this categorical choice.

## S3. Stimulus images with top beta scores per ROI

Table S2 includes lists of stimulus images from the THINGS image set that received the twelve highest beta scores in the single voxel with the highest noise ceiling in a given ROI for each subject.

| ROI | sub-01 | sub-02 | sub-03 | sub-06 |
|---|---|---|---|---|
| FFA | dog_05s | koala_03s | groundhog_04s | face_01b |
| | ferret_03s | undershirt_03s | koala_05s | shower_cap_03s |
| | man_04s | gondola_03s | ferret_05s | rat_03s |
| | skunk_06s | sweatsuit_01s | tiger_03s | girl_04s |
| | weasel_02s | sloth_06s | face_03s | lion_01b |
| | warthog_01b | boy_03s | face_01b | boy_04s |
| | beaver_02s | lion_06s | warthog_04n | woman_03s |
| | squirrel_02s | cufflink_04s | weasel_05n | hamster_01b |
| | gargoyle_03s | trap_04s | football_helmet_02s | man_04s |
| | chess_piece_06s | chalice_04s | bear_01b | otter_06n |
| | gargoyle_02s | possum_03s | wig_01s | boy_01b |
| | lion_01b | poster_04s | hamster_05s | snorkel_05s |
| EBA | hip_02s | chin_04s | hula_hoop_06s | sweatsuit_05s |
| | sweatsuit_05s | tuxedo_01b | gondola_03s | hula_hoop_04s |
| | hip_01b | monkey_02s | scarecrow_02s | penguin_06s |
| | overalls_03s | koala_06s | lawnmower_01b | hula_hoop_03s |
| | kangaroo_05s | leggings_04s | uniform_03s | coverall_05s |
| | subway_01b | sloth_04s | clarinet_02n | leopard_04s |
| | wheelbarrow_05s | footbath_06s | sloth_03s | elephant_03n |
| | boa_02s | undershirt_03s | clarinet_06s | sweatsuit_03s |
| | uniform_02s | chest1_04s | elephant_03n | horse_04s |
| | lawnmower_01b | handcuff_01b | uniform_04s | pogo_stick_05s |
| | raft_01b | shower_cap_05s | leggings_04s | wolf_03s |
| | tuxedo_05s | sweatsuit_01s | coat_06s | pogo_stick_01s |
| PPA | dishwasher_06s | windowsill_02s | guardrail_03s | computer$_{s}creen$_05$s$ |
| | sink_05s | television_04s | guillotine_06s | fence_03s |
| | chicken_wire_02s | stair_01b | stair_01b | canoe_03s |
| | burner_03s | towel_rack_04s | tent_04s | windowsill_05s |
| | locker_05s | stair_05s | railing_03s | train_03s |
| | punching_bag_01b | scanner_05s | mailbox_03s | hedge_01b |
| | cassette_03s | projector_04s | windowsill_03s | candelabra_05s |
| | anchor_05s | guardrail_06s | mosquito_net_01b | bench_03s |
| | fence_02s | anvil_03s | stair_06s | stopwatch_01b |
| | bunkbed_05s | doormat_01b | windowsill_06s | sink_04s |
| | trashcan_06s | drawer_02s | guardrail_01b | rug_04s |
| | birdcage_03s | tray_04s | fence_02s | shower_03s |

**Table S2. Top-12 stimulus images per ROI .**

## S4. Mean reaction times

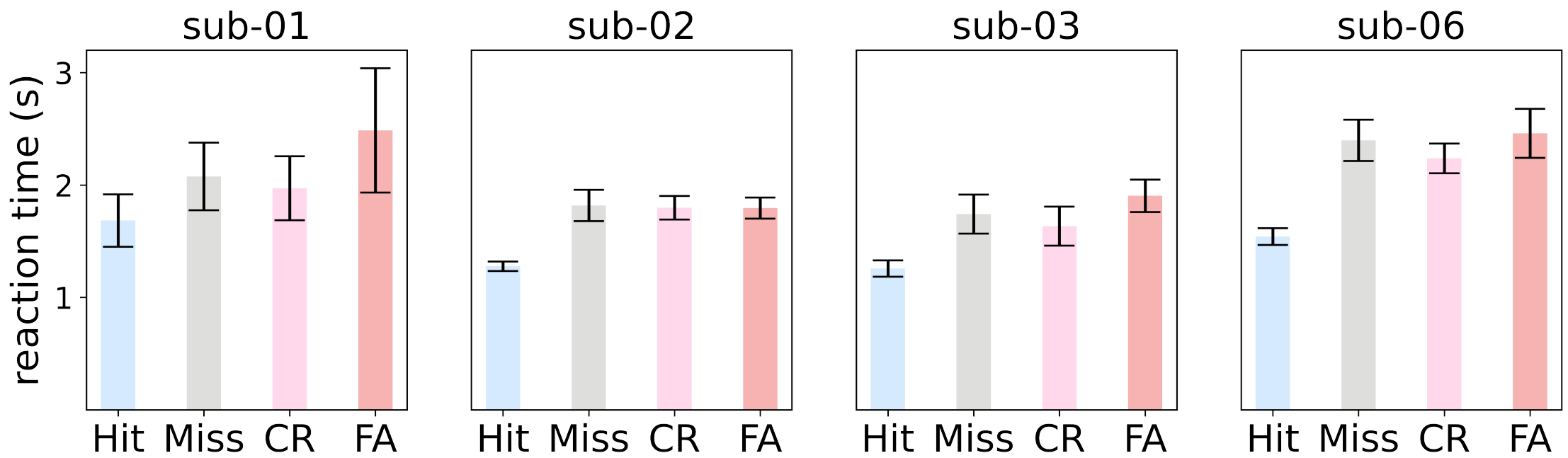

**Fig. S1. Reaction time per type of response.** Reaction time in seconds for hits, misses, correct rejections (CR) and false alarms (FA) is averaged across sessions for each subject. Hit and Miss reaction times are shown separately for within-session (pale blue) and between-session (darker blue) repetitions. Responses error bars represent the standard deviation.

## S5. Dimensionality reduction analyses in functionally defined ROIs (t-SNE)

We conducted data-driven dimensionality reduction analyses to visualize the representation of semantic information in functionally defined visual brain regions. Specifically, we generated t-distributed stochastic neighbor embedding (t-SNE) plots (1) from beta scores estimated with GLMsingle (section 2.7) within subject-specific ROIs delineated according to the following criteria:

- **Face-specific ROIs:** concatenation of voxels from the FFA and OFA. Final ROI extents: 409 voxels (sub-01), 327 voxels (sub-02), and 501 voxels (sub-03).
- **Place-specific ROIs:** concatenation of voxels from the PPA, OPA and MPA. Final extents: 176 voxels (sub-01), 190 voxels (sub-02), and 537 voxels (sub-03).
- **Early visual cortex ROI:** voxels from V1, V2 and V3, identified with Neuropythy (see 2.10). Final extents: 4458 voxels (sub-01), 4543 voxels (sub-02), and 4274 voxels (sub-03).

For this analysis, FFA, OFA, PPA, OPA and MPA boundaries included voxels with the highest ranking t-scores (no lower than 3.72) on the relevant fLoc contrast (face or place > all other conditions, unsmoothed BOLD data) within a group-derived binary parcel (2, 3) (e.g., FFA) warped to the subject's native functional (T1w) space and smoothed with a FWHM=5 kernel. A cut-off was set for the number of selected voxels not to exceed 30% of the group parcel voxel count (post-warping, pre-smoothing).

For each ROI, one t-SNE plot was generated from trial-specific beta scores (excluding no-response trials) and another from image-specific beta scores averaged across repetitions (including only images with three repetitions with recorded responses). Beta scores were z-scored within voxel and thn reduced with PCA, keeping the top 50 components (PCs). T-SNE plots generated with Scikit-learn 1.0.1 were initialized with the betas' first two PCs scaled by 0.0001 of the first PC's standard deviation (perplexity = 100, learning rate = 500, max 2000 iterations).

We used higher-order WordNet categorical labels ("animal", "plant", "vehicle") and object concept ratings between 0 and 7 (i.e., "moves", "size", "natural") from the THINGSplus project (4) to annotate image content within the plots' clusters. Figures S2 (face-specific ROIs), S3 (place-specific ROIs) and S4 (low-level virtual ROIs) showcase annotated t-SNE plots derived from image-wise (top row) and trial-wise (bottom row) beta values for sub-01, sub-02 and sub-03. Clustering patterns indicate the greatest coherence in face-specific ROIs, and the least coherence in low-level visual ROIs.

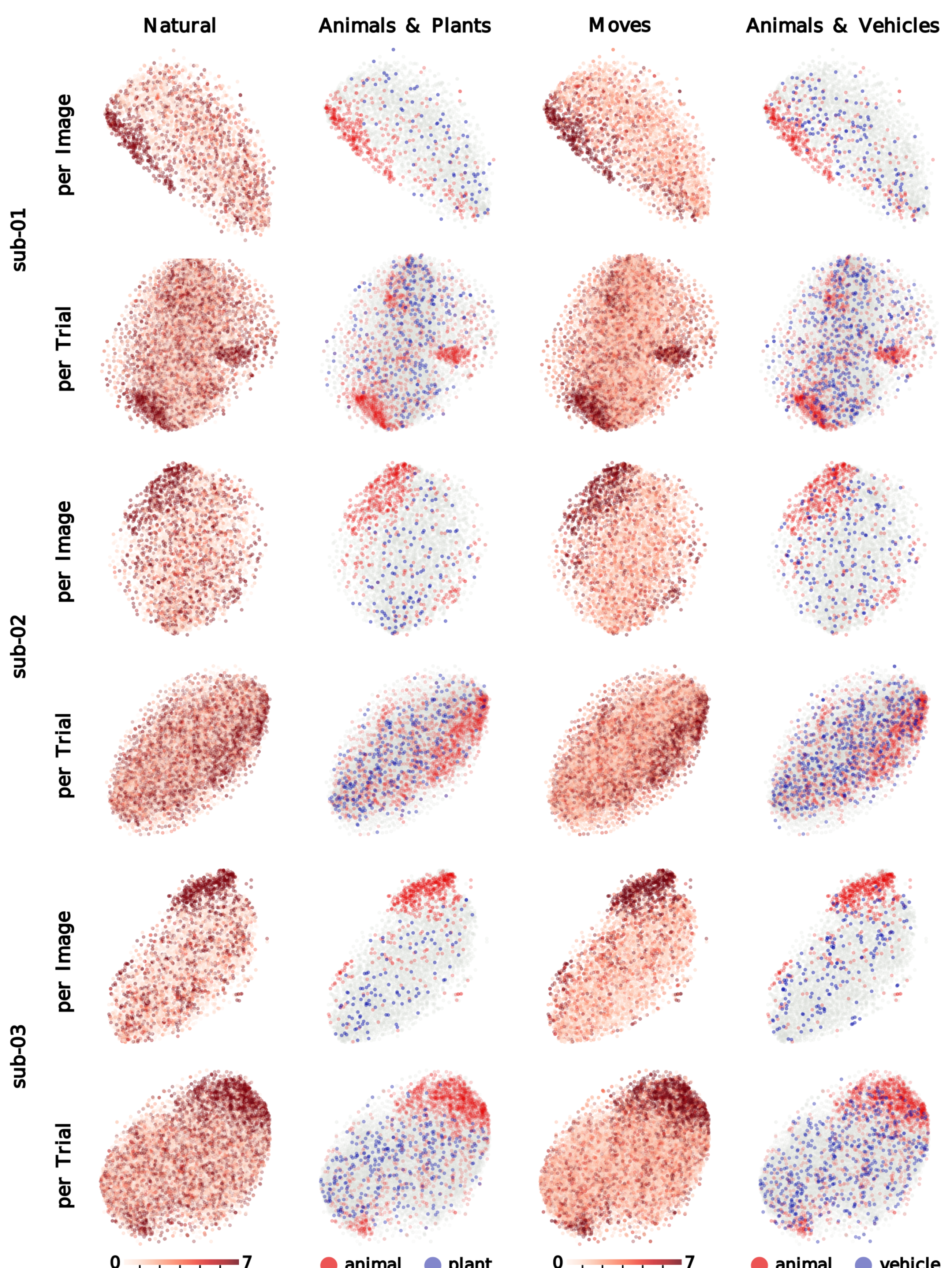

**Fig. S2. Dimension reduction analyses (t-SNE) in face-specific ROIs.** T-SNE plots on 50 principal components derived from normalized beta scores per trial and per image from face-specific ROIs (FFA and OFA) defined functionally for each subject. ROIs included 409, 327 and 501 voxels for sub-01, sub-02 and sub-03, respectively. Image content is annotated with categorical labels (i.e., "animal", "plant", "vehicle") and object concept ratings ranging between 0 and 7 (i.e., "moves", "natural") from the THINGSplus project (4).

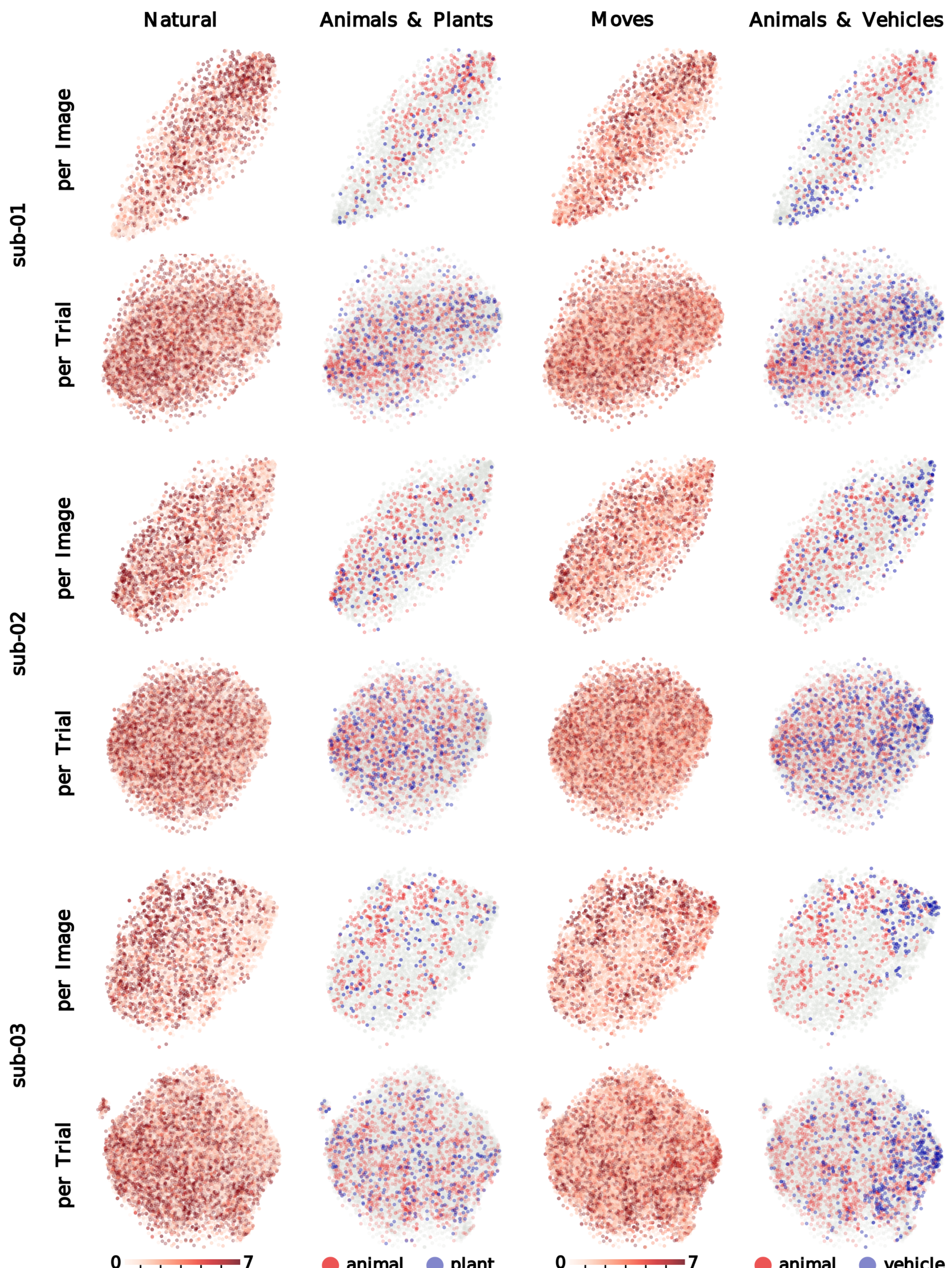

**Fig. S3. Dimension reduction analyses (t-SNE) in scene-specific ROIs.** T-SNE plots on 50 principal components derived from normalized beta scores per trial and per image from scene-specific ROIs (PPA, OPA and MPA) defined functionally for each subject. ROIs included 176, 190 and 537 voxels for sub-01, sub-02 and sub-03, respectively. Image content is annotated with categorical labels (i.e., "animal", "plant", "vehicle") and object concept ratings ranging between 0 and 7 (i.e., "moves", "natural") from the THINGSplus project (4).

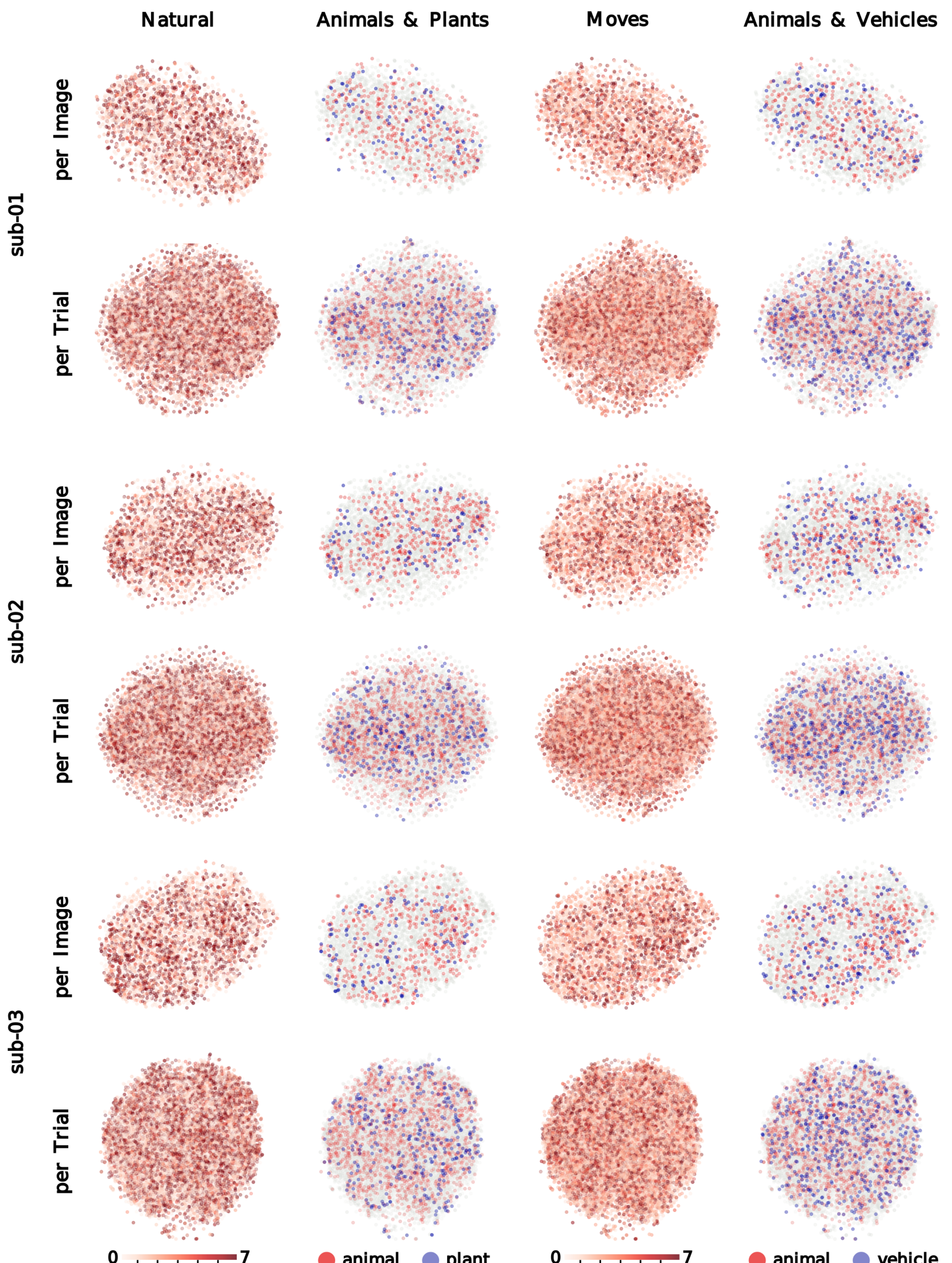

**Fig. S4. Dimension reduction analyses (t-SNE) in low-level visual ROIs.** T-SNE plots on 50 principal components derived from normalized beta scores per trial and per image from low-level visual ROIs (V1, V2 and V3) defined functionally for each subject. ROIs included 4458, 4543 and 4274 voxels for sub-01, sub-02 and sub-03, respectively. Image content is annotated with categorical labels (i.e., “animal”, “plant”, “vehicle”) and object concept ratings ranging between 0 and 7 (i.e., “moves”, “natural”) from the THINGSplus project (4).